 \documentstyle[prl,aps,twocolumn]{revtex}

       \newcommand{\beq}{\begin{equation}}
       \newcommand{\eeq}{\end{equation}} 
       \newcommand{\beqa}{\begin{eqnarray}}
       \newcommand{\eeqa}{\end{eqnarray}}

\def\p{\partial}

\def\s{\hat\sigma}
\def\r{\rho}
\def\k{\kappa}
\def\C{{\cal C}}\def\Q{{\cal Q}}
 \def\half{{\scriptstyle\frac{1}{2}}} 
\def\ihalf{{\scriptstyle\frac{i}{2}}}

\begin{document}

        \draft

        \twocolumn[\hsize\textwidth\columnwidth\hsize\csname
        @twocolumnfalse\endcsname

        \title{Royal Road to Coupling Classical and Quantum Dynamics}
        \author
        {Lajos Di\'osi$^{1,2}$, Nicolas Gisin$^3$, and Walter T. Strunz$^4$}
   
        \address{
        \protect\small\em $^1$Institute for Advanced Study, Wallotstr. 19, 
        D-14193 Berlin, Germany\\ 
        \protect\small\em $^2$ Research Institute for Particle and Nuclear
        Physics, 1525 Budapest 114, POB 49, Hungary\\
        \protect\small\em $^3$Group of Applied Physics, University of Geneva, 
        1211 Geneva 4, Switzerland\\
        \protect\small\em $^4$Fachbereich Physik, Universit\"at GH Essen, 
        45117 Essen, Germany}
        \date{September 2, 1999}

        \maketitle

        \begin{abstract}
        We present a consistent framework of coupled classical and quantum 
        dynamics. Our result allows us to overcome severe limitations of 
        previous phenomenological approaches, like evolutions that do not 
	preserve the positivity of quantum states or that allow to activate 
	quantum nonlocality for superluminal signaling. A `hybrid' 
	quantum-classical density is introduced and its evolution equation 
	derived. The implications and applications of our result are numerous: 
	it incorporates the back-reaction of quantum on classical variables, 
	it resolves fundamental problems encountered in standard mean field 
	theories, and remarkably, also the quantum measurement process, i.e.
	the most controversial example of quantum-classical interaction is 
	consistently described within our approach, leading to a theory of 
	dynamical collapse.
        \end{abstract}


        \vskip2pc]
	\narrowtext

{\it Opinions vary} about the coexistence of and the interaction between 
classical and quantum systems. In orthodox quantum theory, classical 
macrosystems and quantized microsystems coexist, their interaction is 
described asymmetrically. The influence of macrovariables upon microsystems 
is precisely taken into account as external forces. The {\it back-reaction\/} 
of quantized microsystems upon classical macrosystems is largely
ignored, except for {\it detector variables} which are typically sensitive to 
certain microvariables. The theory of this specific back-reaction, 
called measurement theory \cite{Neu}, predicts the statistics of the final
states after the interaction. Yet, the interpretation of quantized dynamics is 
exclusively based on this non-dynamical model of back-reaction (cf. collapse 
of the quantum state), without it we could not test the validity of quantized
dynamics at all. Possibly, quantization extends to macrosystems, indeed
the criteria of being macro- or microscopic are loosely if ever defined.
Contrary to quantized microvariables, quantized macrovariables may have 
significant back-reactions on generic classical macrovariables as well. 
This becomes apparent in the widely used mean-field approximation \cite{Ros} 
which, however, has several fundamental drawbacks \cite{Kib} as we shall recall 
in this Letter. The measurement theory also describes back-reaction, but only 
for idealized detectors. In some attempts to define quantum gravity, matter and
some fields are quantized, while other fields (gravity in particular) are 
treated as classical, requiring thus a definite {\it hybrid} -- i.e. a coupled 
classical and quantum -- dynamics. Thus, a general model of the back-reaction 
is desirable. Such a theory would describe the "collapse" of the wave function
dynamically \cite{SheSud}, would replace mean-field approximations in a 
systematic way \cite{Ale,BouTra,DioHal}, and could have deep implications for 
quantum cosmology. The first conceptual attempts \cite{SheSud,Ale,BouTra} were 
followed by ups and downs \cite{UD}, until severe limitations have been 
clarified \cite{Sal}. In this Letter we use a straightforward transformation 
of the problem which automatically leads to a consistent model of hybrid 
dynamics. 

{\it The difficulties} to overcome in our Letter are unrelated to the
high complexity of the emblematic mean-field equation of quantum cosmology.
Actually, they lie in foundational principles. To illustrate these
difficulties, we assume a quantum Pauli-spin ${\bf{\s}}$ interacting with a
classical harmonic oscillator of Hamiltonian
$H_\C(x,p)=\half(p^2 + x^2)$. The spin interacts with the "magnetic" field of
the oscillator via the Hamiltonian
\beq
\hat H_I=\k\s_3 p.
\label{spinoscill}
\eeq
Regarding $\hat H=H_\C+\hat H_I$ as the total Hamiltonian,
it is natural to prescribe the Heisenberg equation of motion 
\hbox{$\p_t{\bf\hat\sigma}=i[\hat H,{\bf\hat\sigma}]
=i\k p[{\hat\sigma_3,\bf\hat\sigma}]$} to the spin. 
The classical oscillator momentum satisfies the Hamilton equation
\hbox{$\p_t p=-\p_x\hat H=-x$} but the coordinate cannot satisfy 
\hbox{$\p_t x=\p_p\hat H=p+\k\hat\sigma_3$}
since $x$, being a real number, should not evolve into a matrix.
The obvious way out is to replace the operator $\hat\sigma_3$ by its quantum
expectation value, i.e. to apply some mean field approximation. Yet, if taken 
literally, this implies that quantum expectations can be deduced with
arbitrary precision from the measurement of the classical variables $x$ and 
$p$. Hence, the message is that the classical oscillator should, in some
way, inherit quantum fluctuations from the spin. It comes to one's mind that
$x$ and $p$ should be random variables, but not arbitrary ones. As we shall
demonstrate, mathematical consistency imposes that the classical variables
$x$ or $p$ must never take sharp values: The consistent theory assumes an
unremovable coarse-graining \cite{Dio.Two,Dio.Opt}. 

{\it The mathematical issue\/} is the following. The phase space density
$\r_\C(x,p)$ of a classical canonical system $\C$ satisfies the 
Liouville equation of motion:
\beq
\p_t\r_\C(x,p)=\{H_\C (x,p),\r_\C(x,p)\}_P
\label{C}\eeq
where $H_\C(x,p)$ is the Hamilton function and
$\{f,g\}_P=\p_xf\p_pg-\p_pf\p_xg$ stand for the Poisson bracket. On the other 
hand, the density operator $\hat\r_\Q$ of a (canonical, or maybe discrete) 
quantum system evolves according to the von Neumann equation:
\beq
\p_t\hat\r_\Q=-i[\hat H_\Q,\hat\r_\Q]
\label{Q}\eeq
where $\hat H_\Q$ is the Hamilton operator. To introduce interaction between 
$\Q$ and $\C$, we assume the {\it hybrid} Hamiltonian in the following form:
\beq
\hat H(x,p)=\hat  H_\Q +  H_\C(x,p) + \hat H_{I}(x,p)
\label{Hhyb}\eeq
where the interaction term (and thus the total Hamiltonian, too) is
a Hermitian operator for $\Q$, depending on the phase space coordinates 
of $\C$.

{\it The mean-field\/} approach assumes sharp classical
coordinates $x_t,p_t$ at each time and the current quantum  expectation 
value $H_{MF}(x,p;t)=tr[\hat H(x,p)\hat\r_\Q(t)]$ of the Hamiltonian 
is regarded as the effective Hamilton function for the classical 
subsystem $\C$. The coupled evolution equations then take this form:
\beq
\p_t\hat\r_\Q(t)=-i[\hat H(x_t,p_t),\hat\r_\Q(t)]
\label{MFQ}\eeq
\beq
\p_t x_t=\p_p H_{MF}(x_t,p_t),~~~~\p_t p_t=-\p_x H_{MF}(x_t,p_t).
\label{MFC}\eeq
This approach has well-known deficiencies. In particular, it gives no account 
of indeterminacy of the classical states $x,p$ inherited from the quantum
uncertainties of $\hat\r_\Q$. There is thus an essential nonlinearity in the 
mean-field von-Neumann  Eq.~(\ref{MFQ}) which leads to fundamental conflicts
with principles of locality \cite{NGHPA89}. Furthermore, the mean-field
effective Hamilton function $H_{MF}(x,p)$ will never be the proper 
representative of the interaction when quantum uncertainties in 
$\hat\r_\Q$ are large.

A promising {\it conceptual\/} approach \cite{Ale} uses the {\it hybrid
density\/} $\hat\r(x,p)$ to represent the state of the composite system
$\C\times\Q$. If the subsystems are uncorrelated, then the hybrid
density simply factorizes as $\r_\C(x,p)\hat\r_\Q$. In the general case,
the hybrid density $\hat\r(x,p)$ should be an $(x,p)-$dependent nonnegative
operator, satisfying an overall normalization condition
$tr\int\hat\r(x,p)dxdp=1$. One interprets the marginal distribution 
$\r_\C(x,p)\equiv tr\hat\r(x,p)$ 
as the phase space density of the classical subsystem $\C$ while
the density operator $\hat\r_\Q\equiv\int\hat\r(x,p)dxdp$ represents the
unconditional state of the quantum subsystem $\Q$. {\it Conditional quantum
states\/} are natural to introduce at fixed canonical coordinates $(x,p)$ of 
the classical subsystem \cite{Dio.Two}:
\beq
\hat\r_{xp}\equiv\hat\r(x,p)/\r_C(x,p).
\label{rhocond}\eeq
Aleksandrov \cite{Ale} proposed the following evolution equation for the 
hybrid density:
\beqa\label{Aleks}
\p_t\hat\r(x,p)&=&-i[\hat H(x,p),\hat\r(x,p)]+ \\ 
                &+& \half\{\hat H(x,p),\hat\r(x,p)\}_P
                - \half\{\hat\r(x,p),\hat H(x,p)\}_P~.\nonumber
\eeqa
If $\hat H_{I}(x,p)=0$, then this equation splits into the standard 
Eqs.~(\ref{C}) and (\ref{Q}).

Let us test Aleksandrov's equation on the spin-oscillator system
(\ref{spinoscill}). The generic form of the hybrid state is
\beq
\hat\r(x,p)=\half[1+{\vec s}(x,p){\hat{\vec\sigma}}]\r_\C(x,p),~~~
                                           \vert{\vec s}\vert\leq1,
\label{rhybsp}\eeq
where ${\vec s}(x,p)$ is the spin-vector correlated with the oscillator's 
state. We read off the conditional quantum state (\ref{rhocond}) of the spin:
$\hat\r_{xp}=\half[1+{\vec s}(x,p){\hat{\vec\sigma}}]$. For the hybrid state 
(\ref{rhybsp}) and interaction Hamiltonian (\ref{spinoscill}), 
Aleksandrov's Eq.~(\ref{Aleks}) reads:
\beq
\p_t\hat\r=(x\p_p-p\p_x)\hat\r
          -i\k p[\s_3,\hat\r]-\half\k[\s_3,\p_x\hat\r]_{+}~.
\label{Aleksspin}\eeq
This equation easily violates the positivity condition 
$\vert{\vec s}\vert\leq1$ on $\hat\r$. E.g., the initial normalized
polarization vector
\beq\label{svec}
{\vec s}=\frac{1}{x^2+p^2+1}(2x,2p,x^2+p^2-1),
\eeq
leads to $\vert{\vec s}\vert>1$ for all $x>0$ under the evolution 
(\ref{Aleksspin}). So, the naive Eq.~(\ref{Aleks}) is inconsistent since 
it does not guarantee the positivity of the hybrid density $\hat\r(x,p)$ 
\cite{BouTra}. One sees that the mathematical textures of the classical 
$(\C)$ and quantum $(\Q)$ systems, though well understood each separately, 
are not at all trivial to couple.

{\it A royal road} offers nonetheless.
Let us quantize canonically $\C$ as well. We do so temporarily and, at the
end of the day, we regard it classical again. The hybrid Hamiltonian
(\ref{Hhyb}) transforms into the total Hamilton operator of the 
fully quantized system $\C\otimes\Q$:
\beq
:\!\!\hat H(\hat x,\hat p)\!\!:=\hat  H_\Q 
+  :\!\!H_\C(\hat x,\hat p)\!\!: + :\!\!\hat H_{I}(\hat x,\hat p)\!\!: 
\label{Hbig}\eeq
where $:\!\!\dots\!\!:$ stand for normal ordering in terms of the usual 
annihilation and creation operators $(\hat x\pm i\hat p)/\sqrt{2}$, 
respectively. Let the equation of motion for the total system's density 
operator $\hat\r$ be the standard von Neumann one:
\beq
\p_t\hat\r=-i[:\!\!\hat H(\hat x,\hat p)\!\!:,\hat\r]~.
\label{Qbig}\eeq
Our royal road is based on coherent states \cite{COH}. Coherent states 
$\vert x,p\rangle$ are eigenstates of the annihilation operator:
\beq
\left(\hat x+ i\hat p\right)\vert x,p\rangle = 
     \left(x+ip\right)\vert x,p\rangle~~.
\label{cohst}
\eeq
Using Bargmann's convention \cite{COH}, the coherent states satisfy the
following differential relation:
\beqa
\left(\p_x-i\p_p\right)\vert x,p\rangle &=& \left(\hat x- i\hat
p\right)\vert x,p\rangle~~, \label{dercohst}\\
\left(\p_x+i\p_p\right)\vert x,p\rangle&=&0~~.
\label{anal}
\eeqa
The latter relation expresses the fact that the bras $\vert x,p\rangle$ 
are entire analytic functions of the complex canonical variable $x+ip$, a 
crucial fact as we shall see. The coherent states form an overcomplete basis
with normalization
\beq
I=\int\vert x,p\rangle\langle x,p\vert 
  \frac{\exp(-\half (x^2 + p^2))}{2\pi}dxdp.
\label{norm}\eeq
It follows from Eqs.(\ref{cohst},\ref{dercohst}) that
\beqa\label{baseq}
&&:\!\!f\!(\hat x,\hat p)\!\!:\vert x,p\rangle = \\
&&:\!\!f\!\left(\frac{x+ip+\p_x-i\p_p}{2},\frac{p-ix+\p_p+i\p_x}{2}\right)
\!\!:\vert x,p\rangle \nonumber
\eeqa
for an arbitrary normal ordered function of the quantized variables
on the l.h.s. On the r.h.s., the same symbols $:\!\!\dots\!\!:$ mean that
all derivations must be done first!

We apply a projection to the density operator $\hat\r$ of the fully 
quantized system $\C\otimes\Q$ and thus reintroduce the  
{\it hybrid density}: 
\beq
\hat\r(x,p)
\equiv tr\Bigl( (\vert x,p\rangle\langle x,p\vert\otimes I)\hat\r\Bigr)
       \frac{\exp(-\half (x^2 + p^2))}{2\pi}.
\label{proj}\eeq
Indeed, this can formally be considered the hybrid density of the composite
system $\C\times\Q$ as if $\C$ were unquantized (i.e. classical) again. 
This is what we are going to do. We can thus {\it derive} the closed 
equation of motion of the hybrid density (\ref{proj}) from the exact 
von Neumann equation (\ref{Qbig}). Using   the basic relation (\ref{baseq})
and the identity (\ref{anal}), we obtain the desired evolution equation: 
\beqa\label{hybeq}
&&\p_t\hat\r(x,p)= \\
&& -i:\!\!\hat H\left(x+{\p_x+i\p_p\over2},p+{\p_p-i\p_x\over2}\right)\!\!:
\hat\r(x,p) + H.C. \nonumber
\eeqa
This {\it hybrid dynamic equation} is our proposal to couple classical systems 
to quantum ones canonically. Note that the hybrid density $\hat\r(x,p)$ 
incorporates our statistical knowledge of $\Q$'s quantum state, of $\C$'s 
classical state, and of their correlations. If $\hat H_{I}(x,p)=0$, then by
integrating both sides of (\ref{hybeq}) over $x$ and $p$ we get the standard 
von Neumann Eq.~(\ref{Q}), as it should be. The trace of both sides, however,
does {\it not} lead to the standard classical dynamics (\ref{C}). Instead, 
we get (for obvious reasons) the evolution equation of a Husimi-function 
\cite{Hus}. To lowest order in the derivatives, however, this is the classical 
Liouville Eq.~(\ref{C}). Hence classical dynamics is recovered if both the 
Hamilton function and the state distribution change slowly with $x$ and $p$
(see e.g. \cite{Mol} and references therein).
 
{\it Consistency} of the hybrid equation of motion (\ref{hybeq}) is, contrary 
to the case of the naive Eq.~(\ref{Aleks}), assured by construction. 
It preserves the positivity of the hybrid density $\hat\r(x,p)$ along with 
a certain analyticity property. In fact, the projection (\ref{proj})
leads always to hybrid densities of the form 
\beq   
\hat\r(x,p)=\frac{\exp(-\half (x^2 + p^2))}{2\pi}
\sum_n\varphi_n(x-ip)\varphi_n^\dagger(x+ip)
\label{ranal}\eeq
where $\varphi_n(x-ip)$ are unnormalized non-orthogonal state vectors for
$\Q$, being complex entire functions of the combinations $x-ip$ of 
$\C$'s classical variables. This positive form is then preserved by our
hybrid equation (\ref{hybeq}). The analyticity condition (\ref{ranal}) 
restricts the possible hybrid states: sharp values of $x,p$ and, generally,
characteristic phase-space dependences inside single Planck cells are 
excluded.  

In particular, the "pure state" form of (\ref{ranal}), i.e. with a single 
$\varphi\varphi^\dagger$ dyadic term, is also preserved. The hybrid 
Schr\"odinger equation of the hybrid state vector $\varphi(x-ip)$ follows from 
(\ref{hybeq}) \cite{QSD}. For completeness, we mention that the polarization 
(\ref{svec}) can be reproduced by 
$\varphi(x-ip)=[(x-ip)\vert+\rangle + \vert-\rangle]/\sqrt{3}$ 
where $\s_3\vert\pm\rangle=\pm\vert\pm\rangle$. 

{\it A post-mean-field approximation\/}, incorporating some of the 
fluctuations of system $\Q$ into its back-reaction on the system $\C$,
is worth to derive. Consider the $1st$-order expansion of the 
Eq.~(\ref{hybeq}) in the derivatives of $\hat H(x,p)$,
\beqa
\p_t\hat\r(x,p)=-i[\hat H(x,p),\hat\r(x,p)]+\nonumber\\
                + \half\{\hat H(x,p),\hat\r(x,p)\}_P
                - \half\{\hat\r(x,p),\hat H(x,p)\}_P\nonumber\\
                 -\ihalf[\p_x\hat H(x,p),\p_x\hat\r(x,p)]
                 -\ihalf[\p_p\hat H(x,p),\p_p\hat\r(x,p)]~.
\label{hybeq1}\eeqa
This equation has two additional terms with respect to the (mathematically
inconsistent) Eq.~(\ref{Aleks}). These two additional terms reduce the domain 
of inconsistency. More importantly, it can be shown that the above equation 
is equivalent to the exact Eq.~(\ref{hybeq}) if $\C$ is harmonic and its 
coupling to $\Q$ is linear in $x$ and $p$. In particular, Eq.~(\ref{hybeq1})
gives a mathematically consistent theory of fully quantized atomic systems 
$(\Q)$ interacting with the fully developed {\it classical} radiation field 
$(\C)$. Moreover, its physics is equivalent with the true fully quantized 
radiation theory \cite{Dio.Opt}.

By taking the trace of Eq.~(\ref{hybeq1}) one can show that the evolution of 
the classical states is a flow:
\beq
\p_t x= \langle \p_p\hat H(x,p) \rangle_{xp},~~
\p_t p=-\langle \p_x\hat H(x,p) \rangle_{xp},
\label{xpflow}\eeq
where $\langle\dots\rangle_{xp}$ stands for the expectation values in the
current conditional quantum state $\hat\r_{xp}$. Obviously this flow resembles 
locally the naive mean-field Eq.~(\ref{MFC}). Here, however, the classical 
state is randomly distributed: it inherits the quantum fluctuations of $\Q$. 
On the other hand, the deterministic evolution of the state's distribution is 
a remarkable fact with significant consequences being discussed elsewhere.

{\it Dynamical collapse\/} of the quantum state is encoded in our 
hybrid evolution equation (\ref{hybeq}). It is a most spectacular feature.
The "presence" of standard collapse will be demonstrated on the 
spin-oscillator model. For the hybrid state (\ref{rhybsp}), our evolution 
equation (\ref{hybeq}) reads:
\beqa 
\p_t\hat\r=(x\p_p-p\p_x)\hat\r
           &-&i\k p[\s_3,\hat\r]-\half\k[\s_3,\p_x\hat\r]_{+}\nonumber\\
           &-&\ihalf\k[\s_3,\p_p\hat\r],
\label{hybeqspin}\eeqa
which differs from the naive Eq.~(\ref{Aleksspin}) by the presence
of the $4th$ term on the r.h.s. This term guarantees the positivity of
the hybrid state $\hat\r(x,p)$ for all times. 
The oscillator plays the role of the Stern-Gerlach apparatus detecting the 
quantized spin-component $\s_3$. The pointer variable of the apparatus
is $x$, it is set initially to zero with precision $\Delta=1$. 
Accordingly, we choose $\r_\C^0(x,p)=(2\pi)^{-1}\exp(-\half (x^2 + p^2))$ 
for the oscillator's initial state. We shall switch on the interaction 
Hamiltonian at $t=0$ for a short time $\tau$ compared to the oscillator 
period, keeping the effective coupling as strong as $g=\k\tau\gg 1$. 
Actually, we replace $\k$ by $g\delta(t)$ \cite{Neu}. 
Let us assume that the spin's initial state is the superposition
$\vert\psi\rangle=c_{+}\vert+\rangle+c_{-}\vert-\rangle$
and the initial hybrid state (\ref{rhybsp}) is the uncorrelated
$
\vert\psi\rangle\langle\psi\vert\r_\C^0(x,p)
$.
Immediately after the interaction the hybrid state $\hat\r(x,p)$ becomes
\beqa
&& \vert c_{+}\vert^2\vert+\rangle\langle+\vert\r_\C^0(x+g,p)\nonumber 
+\vert c_{-}\vert^2\vert-\rangle\langle-\vert\r_\C^0(x-g,p)+\\
&& c_{+}c_{-}^\star\vert+\rangle\langle-\vert \exp(-\half g^2 -
igp)\r_\C^0(x,p) + H.C.~~.
\label{hybout}\eeqa
The trace yields the pointer's state distribution:
\beq
\r_\C(x,p)= \vert c_{+}\vert^2\r_\C^0(x-g,p) 
           +\vert c_{-}\vert^2\r_\C^0(x+g,p)~~.
\label{roscout}\eeq
Since $g \gg 1$, we shall ignore the overlap between the two terms
on the r.h.s., so we can say that the pointer $x$ has swung out to
$g\pm\Delta$ or to $-(g\pm\Delta)$ with the probabilities predicted by the 
standard measurement theory. Invoking Eq.~(\ref{rhocond}), we can easily 
read out the conditional quantum state of the spin from Eq.~({\ref{hybout}) 
[since the off-diagonal terms are damped by $exp(-\half g^2)$, 
we ignore them]: 
\beq
\hat\r_{xp}= \vert\pm\rangle\langle\pm\vert,~~x\approx \pm g,~~p\approx 0.
\label{rspout}\eeq
This shows the standard collapse of the spin's quantum state: the quantum
state is correlated with the classical pointer's position.

{\it Discussing\/} this Letter's results 
we repeat that we are aware of the ambiguous contemporary views
concerning the concept of genuine hybrid systems. Nevertheless, the old
Copenhagen interpretation as well as recent quantum gravity and quantum 
cosmological models assume such hybrid systems. We made the necessary 
compromises to neutralize strict no-go theorems. Our equation (\ref{hybeq}) 
is a first example of hybrid dynamics which is both mathematically 
consistent and physically relevant. The applications of our equation are 
numerous, for example as phenomenological models whenever the mean-field
approximation is poor. Moreover, we derive the post-mean-field
equations which describe back-reaction of quantum fluctuations to first 
order. On the foundational level, we point out that our hybrid dynamics 
(\ref{hybeq}) reproduces the ideal quantum measurement, including the 
collapse of the wave function {\it and} the motion of the classical pointer. 
Let us also stress the close connection to current phenomenological theories
of dynamic collapse which {\it follow} from our hybrid dynamics \cite{QSD}.
Our hybrid theory is likely to be an integrating concept for treating quantum
measurement dynamically and to overcome the inconsistent mean-field method
in quantum cosmology.

We are grateful to Jonathan Halliwell, Ian C Percival and Ting Yu for useful
conversations. LD was partly supported by EPSRC and by OTKA T016047,  
NG by the Swiss National Science Foundation, and
WTS by the Deutsche Forschungsgemeinschaft through the SFB 237
``Unordnung und gro{\ss}e Fluktuationen''.

\end{document}